\documentclass[prb,amsmath,amssymb,superscriptaddress,twocolumn,showpacs]{revtex4}
\usepackage{graphicx}
\usepackage{subfigure}
\usepackage{bm}
\usepackage{color}
\usepackage{float}

\DeclareMathAlphabet{\mathpzc}{OT1}{pzc}{m}{it} 

\definecolor{CommentRed}{rgb}{0.9,0,0}

\begin{document}

\title{Kekule versus hidden superconducting order in graphene-like systems: Competition and coexistence}
\author{Flore K. Kunst}
\affiliation{Dahlem Center for Complex Quantum Systems and Institut f\" ur Theoretische Physik, Freie Universit\" at Berlin, Arnimallee 14, 14195 Berlin, Germany}
\affiliation{Institute for Theoretical Physics, Centre for Extreme Matter and Emergent Phenomena, Utrecht University, Leuvenlaan 4, 3584 CE Utrecht, The Netherlands}
\author{Christophe Delerue}
\affiliation{IEMN, Department of ISEN, UMR CNRS 8520, 59046 Lille, France}
\author{Cristiane Morais Smith}
\affiliation{Institute for Theoretical Physics, Centre for Extreme Matter and Emergent Phenomena, Utrecht University, Leuvenlaan 4, 3584 CE Utrecht, The Netherlands}
\author{Vladimir Juri\v{c}i\'{c}}
\affiliation{Institute for Theoretical Physics, Centre for Extreme Matter and Emergent Phenomena, Utrecht University, Leuvenlaan 4, 3584 CE Utrecht, The Netherlands}

\begin{abstract}
 We theoretically study the competition between two possible exotic superconducting orders that may occur in graphene-like systems, assuming dominant nearest-neighbor attraction: the gapless hidden superconducting order, which renormalizes the Fermi velocity, and the Kekule order, which opens a superconducting gap. We perform an analysis within the mean-field theory for Dirac electrons, at finite-temperature and finite chemical potential, as well as at half filling and zero-temperature, first excluding the possibility of the coexistence of the two orders. In that case, we find the dependence of the critical (more precisely, crossover) temperature and the critical interaction on the chemical potential. As a result of this analysis, we find that the Kekule order is preferred over the hidden order at both finite temperature and finite chemical potential. However, when the coexistence of the two superconducting orders is allowed, using the coupled mean-field gap equations, we find that above a critical value of the attractive interaction a mixed phase sets in, in  which these orders coexist. We show that the critical value of the interaction for this transition is greater than the critical coupling for the hidden superconducting state in the absence of the Kekule order, implying that there is a region in the phase diagram where the Kekule order is favored as a result of the competition with the hidden superconducting order. The hidden superconducting order, however, eventually sets in and coexists with the Kekule state. According to our mean-field calculations, the transition from the Kekule to the mixed phase is of the second order, but it may become first order when fluctuations are considered. Finally, we investigate whether these phases could be possible in honeycomb superlattices of self-assembled semiconducting nanocrystals, which have been recently experimentally realized with CdSe and PbSe.
\end{abstract}


\maketitle

\section{Introduction}

Ever since its isolation, graphene and graphene-related topics have attracted much attention from both theoretical and experimental condensed-matter communities.\cite{geim} The monolayer of carbon atoms is light, transparent, flexible, strong and conductive, which makes it a perfect candidate for industrial applications. The appearance of these properties in the material are for a large part due to the assembly of the atoms in a honeycomb geometry, which consists of two interpenetrating triangular Bravais lattices, referred to as sublattices $A$ and $B$. The lattice symmetry together with time-reversal gives rise to the hallmark feature of the graphene system - pseudo-relativistic massless Dirac fermions, which are low-energy quasiparticles close to the Dirac points located at the two inequivalent, time-reversal conjugate ${\bf K}$ and ${\bf K}'=-{\bf K}$ momenta at the corners of the Brillouin zone (BZ). As a consequence, the density of states linearly vanishes close to the Dirac points. The semimetallic ground state is therefore protected against the effects of weak interactions, and an intrinsic superconducting state in half-filed graphene could thus only be possible for sufficiently strong attractive interactions.\cite{HJR} Onsite attraction supports an $s-$wave spin-singlet superconducting state,\cite{zhao,uchoa-2005} whereas nearest-neighbor attraction may lead to the formation of the Kekule state, which breaks translational lattice symmetry and opens up a gap at the Dirac points,\cite{roy} while, at the same time, a gapless superconductor may also set in.\cite{uchoa} Therefore, it is of fundamental importance to address the competition of these two superconducting orders, and this is precisely the aim of the present paper. This problem is also important in light of the recent progress in inducing superconductivity in graphene via the proximity effect, by building a Josephson junction,\cite{heersche} as well as by growing a graphene sheet on rhenium.\cite{tonnoir}

The fact that superconductivity is not an intrinsic property of graphene has led to the search for the Dirac superconductor in materials with graphene-like properties. A recently proposed type of engineered Dirac material consists of semiconducting nanocrystals with a truncated-cubic shape that self-assemble into a honeycomb superlattice.\cite{kalesaki} The motivation to build these materials was to study the electronic band structures that emerge when gapped, semiconducting systems are combined with features similar to those of graphene, by arranging the nanocrystals in a honeycomb lattice. These materials have been experimentally realized for rocksalt PbSe and zinc-blende CdSe nanocrystals, which form honeycomb superlattices with a lattice parameter $a\simeq6$nm via the attachment of the $\{100\}$ facet of the nanocrystal.\cite{bone} The electronic band structure has been theoretically described for PbSe, CdSe, and HgTe superlattices, and it has been predicted that they exhibit Dirac cones in the conduction band above a wide gap enlarged by the quantum confinement.\cite{kalesaki, crispaper, 2Dmat}

Here, we investigate whether phonon-driven superconductivity would be possible in artificial graphene samples made of PbSe or CdSe nanocrystals. We consider semiconducting sheets that are either free-standing or capped with LiF, a dielectric which has been recently used to isolate and passivate nanocrystal layers.\cite{Tang10,Choi13} All of these systems are treated through an effective model where each nanocrystal is modelled as a superatom with a single effective $s$ orbital representing the lowest conduction state characterized by an $s$-envelope wavefunction.
\cite{kalesaki, crispaper, 2Dmat} We consider only the electrons close to the Dirac point. Furthermore, we assume that the electrons couple to a single Einstein phonon per superatom site, which corresponds to the longitudinal optical (LO) phonon for CdSe, PbSe, or LiF. This particular model is considered because it can account for the features observed in scanning tunneling spectroscopy experiments on CdSe\cite{sun} or PbSe\cite{Overgaag09} nanocrystals.

In the problem studied in this paper, the LO phonon couples to the effective $s$-electrons on the same site as well as on nearest-neighbor sites. The electrons are described by a tight-binding Hamiltonian, where the electron-phonon coupling includes both on-site and nearest-neighbor terms. We then integrate out the phonons to derive the effective electron-electron interaction. An estimate for this effective interaction is obtained based on a numerical analysis, and we find that for both PbSe and CdSe the effective interaction is repulsive, but can become attractive when the superlattice is capped by LiF. The renormalized values for the attractive on-site and nearest-neighbor interactions that we find indicate that in these materials only on-site pairing should occur. Nevertheless, we theoretically investigate the more intricate case when nearest-neighbor interactions dominate over the on-site one, with the aim of motivating further experimental search for graphene-like materials that could fulfill these conditions and exhibit the elusive Kekul\'e or hidden superconducting order described here.

We consider both on-site and nearest-neighbor pairings, and the electron-electron interaction is then decoupled in these channels using mean-field approximation.
 In this setup, we consider the problem of the competition of the Kekule and the hidden supeconductors at both finite temperature and finite chemical potential in the vicinity of the Dirac points, and derive the gap equations for these order parameters. First, excluding the possibility of the coexistence of the two orders, based on these equations, we find the dependence of the critical (more precisely, crossover) temperature and the critical interaction on the chemical potential, and analytical solutions are obtained in the quantum-critical (strong coupling) and the BCS (weak coupling) limits, for both the Kekule and the hidden order. According to our results, the Kekule order is preferred over the hidden order at both finite temperature and finite chemical potential. Second, when we allow for the possibility of coexistence, based on self-consistent mean-field gap equations, we obtain that above a critical value of the attractive interaction a mixed phase sets in where the two superconducting orders in fact do coexist. We show that the critical value for its onset is greater than the critical coupling for the hidden superconducting state in absence of the Kekule order. Therefore, there is a region in the phase diagram where the Kekule order is favored when competing with the hidden superconducting state. However, the latter eventually sets in and coexists with the Kekule superconductor. Finally, according to our mean-field calculations, the transition from the Kekule to the mixed phase is of the second order, but this result may change when fluctuations are taken into account.

The paper is organized as follows. In Sec.~\ref{section 2}, we introduce the model that describes electrons, phonons, and their interaction in the system. In Sec.~\ref{section 3}, we define s-wave, Kekule and hidden superconducting order parameters, and obtain the corresponding mean-field Hamiltonians. We derive and analyze the gap equations for Kekule and hidden orders in Sec.~\ref{section 4}. Results concerning the effective attractive interactions in self-assembled nano-crystals and the conclusions are presented in Sec.~\ref{section 5}. Calculational details are presented in the Appendices.

\section{Model} \label{section 2}

In this Section, we derive the effective model to study the superconducting properties of the system. We introduce full tight-binding and electron-phonon Hamiltonians, after which we integrate out the phonon modes to obtain the effective model.

\subsection{Tight-binding model}

We describe the system using a tight-binding Hamiltonian $H_{\rm full}$, which includes nearest-neighbor hopping, Hubbard terms for electron-pairing on-site and between nearest-neighbors, the chemical potential, and electron-phonon coupling
\begin{equation}
H_{\rm full} = H_{\rm Hub} + H_\mu + H_{\rm el-ph}. \label{eqn fullhamiltonianstatedinbegining}
\end{equation}
The Hubbard Hamiltonian reads
\begin{eqnarray}
&& H_{\rm Hub} = -t \sum_{\left<i,j\right>, \sigma} a^{\dagger}_{i, \sigma} b_{j, \sigma} + h.c. \nonumber \\
&& + U \sum_i a^{\dagger}_{i, \uparrow} a^{\dagger}_{i, \downarrow} a_{i, \downarrow} a_{i, \uparrow} + a \rightarrow b \nonumber \\
&& + V \sum_{\left<i,j\right>; \sigma, \sigma'} a^{\dagger}_{i, \sigma} a_{i, \sigma} b^{\dagger}_{j, \sigma'} b_{j, \sigma'}, \label{eqn hubbardhamiltonianinrealspace}
\end{eqnarray}
where $\left<i,j\right>$ denotes nearest-neighbor sites $i$ and $j$, $a^{\dagger}_{i, \sigma}$ ($a_{i, \sigma}$) are creation (annihilation) operators for an electron on sublattice $A$ at site $i$ with spin $\sigma$, $t$ is the nearest-neighbor hopping parameter, and $U$ ($V$) are the on-site (nearest-neighbor) Coulomb interactions, respectively. The hopping parameter $t$ depends on the size and the shape of the nanocrystals, as well as on the number of atoms connecting neighboring sites. Typically, this parameter is of the order of $10$ meV.\cite{kalesaki, crispaper, 2Dmat} The Hamiltonian for the chemical potential reads

\begin{equation}
H_{\mu} = -\mu \sum_{i, \sigma} \left( a^{\dagger}_{i, \sigma} a_{i, \sigma} + b^{\dagger}_{i, \sigma} b_{i, \sigma} \right), \label{eqn reciprocalspacehamiltonianforthechemicalpotential}
\end{equation}
where $\mu \lesssim 0.5 \, t$ to ensure the validity of the Dirac description of the electrons.
Finally, the electron-phonon Hamiltonian has the form
\begin{eqnarray}
&& H_{\rm el-ph} = \hbar \omega_{E} \sum_{i} c^{\dagger}_{A, i} c_{A, i} \nonumber \\
&& + V_0 \sum_{i, j; \sigma} \delta_{{\bf r}_i, {\bf r}_j} a^{\dagger}_{i, \sigma} a_{i, \sigma} \left(c^{\dagger}_{A, j} + c_{A, j}\right) \nonumber \\
&& + \tilde{V}_0 \sum_{i, j; \sigma; \alpha} \delta_{{\bf r}_i, {\bf r}_j - \boldsymbol\delta_{\alpha}} b^{\dagger}_{i, \sigma} b_{i, \sigma} \left(c^{\dagger}_{A, j} + c_{A, j}\right) \nonumber \\
&& + A \leftrightarrow B, \label{eqn electronphononhamiltonianrealspace}
\end{eqnarray}
where the phonon frequency for phonons on sublattices $A$ and $B$ is equal, $c^{\dagger}_{A,i}$ ($c_{A,i}$) create (annihilate) a phonon on sublattice $A$ at site $i$, $\boldsymbol\delta_\alpha$ connects nearest-neighbor sites with $\alpha= 1, \, 2, \, 3$, and $V_0$ and $\tilde{V}_0$ are the coupling constants.

\subsection{Effective model}

We now integrate out the phonons in Eq.~(\ref{eqn electronphononhamiltonianrealspace}), which results in additional contributions to the Hubbard terms $U$ and $V$ in Eq.~(\ref{eqn hubbardhamiltonianinrealspace}). To this end, we use
\begin{equation}
\mathcal{Z} = \int \mathcal{D}\left[\psi^\dagger, \psi \right] \int \mathcal{D}\left[\phi^\dagger, \phi \right] {\rm e}^{-\frac{1}{\hbar \beta} S \left[\psi^\dagger, \psi; \phi^\dagger, \phi \right]}, \nonumber
\end{equation}
where $\phi$ ($\psi$) is an phonon (electron) field, the inverse temperature $\beta = (k_B T)^{-1}$, and the action is given by
\begin{eqnarray}
S \left[\psi^\dagger, \psi; \phi^{\dagger}, \phi\right] &=& \int_0^{\hbar \beta} {\rm d} \tau \left[\psi^{\dagger}(\tau) \partial_\tau \psi (\tau) + \phi^{\dagger}(\tau) \partial_\tau \phi (\tau) \right. \nonumber \\
&& \left. + H\left(\psi^\dagger, \psi; \phi^{\dagger}, \phi \right) \right]. \nonumber
\end{eqnarray}

The electron-phonon action for the phonon on sublattice $A$ yields
\begin{eqnarray}
&& S_{{\rm el-ph}, A} \left[\psi^\dagger, \psi; \phi^\dagger, \phi\right] = \sum_{{\bf q}, n} \phi^\dagger_{A, {\bf q}, n} \left(-i \hat{\omega}_n + \hbar \omega_E \right) \phi_{A, {\bf q}, n} \nonumber \\
&& + \frac{1}{\sqrt{\hbar \beta N}} \sum_{{\bf q}, \sigma, n} \left(V_0 \,\rho_{A, {\bf q}, \sigma, n} + \tilde{V}_0\, \gamma_{{\bf q}}\, \rho_{B, {\bf q}, \sigma, n} \right) \nonumber \\
&& \times \left( \phi^\dagger_{A, - {\bf q}, -n} +\phi_{A, {\bf q}, n} \right), \label{eqn electronphononactionforsublatticea}
\end{eqnarray}
where $\hat{\omega}_n = 2n \pi/ (\hbar \beta)$ for $n \in \mathbb{Z}$ is the Matsubara frequency for bosons,  $\gamma_{{\bf k}} \equiv \sum_{\alpha} {\rm e}^{i {\bf k} \cdot \boldsymbol\delta_\alpha}$, $2N$ is the number of atoms in the system, and $\rho_{A, {\bf q}, \sigma, n}$ is the  electron density
\begin{equation}
\rho_{A, {\bf q}, \sigma, n} \equiv \sum_{{\bf k}, m} \psi^{\dagger}_{A, {\bf k} + {\bf q}, \sigma, m + n} \psi_{A, {\bf k}, \sigma, m}. \nonumber
\end{equation}
The electron-phonon action for a phonon on sublattice $B$ is obtained by the substitution $A \rightarrow B$ in the above equation. Completing the square and integrating out the phonon fields  leads to the Hubbard terms $U$ and $V$ in Eq.\ (\ref{eqn hubbardhamiltonianinrealspace}) renormalized by the electron-phonon coupling
\begin{eqnarray}
&& S_{{\rm eff}, U, V} \left[\psi^\dagger, \psi\right] = - \sum_{{\bf q}, n} \left[\tilde{U}({\bf q}) \rho_{A, {\bf q}, \downarrow, n} \rho_{A, -{\bf q}, \uparrow, -n} + A \rightarrow B\right] \nonumber \\
&& - \sum_{{\bf q}, n} \sum_{\sigma, \sigma'} \tilde{V}({\bf q}) \rho_{A, {\bf q}, \sigma, n} \rho_{B, -{\bf q}, \sigma', -n}, \label{eqn effectivehubbardtildeutildevaction}
\end{eqnarray}
where
\begin{eqnarray}
&& \tilde{U}({\bf q}) = - \frac{1}{\hbar \beta} \frac{1}{N} \left[U - \frac{2}{\hbar \omega_E} \left(V_0^2 + 9 \tilde{V}_0^2 \right) \right], \label{eqn effectivehubbarduinteraction} \\
&& \tilde{V}({\bf q}) = - \frac{1}{\hbar \beta} \frac{1}{N} \left[3 V - \frac{12}{\hbar \omega_E} V_0 \tilde{V}_0 \right]. \label{eqn effectivehubbardvinteraction}
\end{eqnarray}
The details of the derivation are presented in App.~\ref{appendixsection integratingoutphonons}. The effective Hubbard $\tilde{U}({\bf q})$ and $\tilde{V}({\bf q})$ are defined with a minus sign in the prefactor implying that for an attractive interaction these terms are positive.

\section{Mean-Field Approximation} \label{section 3}

Motivated by the possibility that the effective Hubbard interactions may turn out to be attractive, we consider the superconducting instabilities in the artificial graphene samples.
We first define superconducting order parameters, and then use the mean-field approximation to decouple the electron-electron interaction in Eq.~(\ref{eqn effectivehubbardtildeutildevaction}).

\subsection{Order Parameters}

The Hubbard Hamiltonian reads
\begin{eqnarray}
&& H_{{\rm eff}, U, V} = - \tilde{U} \sum_i a^\dagger_{i, \downarrow} a_{i, \downarrow} a^\dagger_{i, \uparrow} a_{i, \uparrow} + a \rightarrow b \nonumber \\
&& - \tilde{V} \sum_{\left<i,j\right>} \sum_{\sigma, \sigma'} a^\dagger_{i, \sigma} b^\dagger_{j, \sigma'} b_{j, \sigma'} a_{i, \sigma}, \label{eqn realspacehamiltonianwithtildeuandtildev}
\end{eqnarray}
where it is assumed from now on that both $\tilde{U}$ and $\tilde{V}$ are positive. The order parameters corresponding to the on-site and nearest-neighbor pairing, respectively, have the following form
\begin{eqnarray}
&& \Delta_0 = \left<a_{i, \downarrow} a_{i, \uparrow} \right> = \left<b_{i, \downarrow} b_{i, \uparrow} \right>, \\
&& \Delta_{\sigma', \sigma} \left({\bf r}_i, {\bf r}_j \right) = \left<b_{j, \sigma'} a_{i, \sigma} \right>,
\end{eqnarray}
where $\Delta_0$ represents the standard $s$-wave order parameter, and a general form for the nearest-neighbor order parameter $\Delta_{\sigma', \sigma} \left({\bf r}_i, {\bf r}_j \right)$ is assumed.\cite{roy}

The electron densities in Eq.~(\ref{eqn realspacehamiltonianwithtildeuandtildev}) can be decoupled via a mean-field approximation
\begin{eqnarray}
&& a_{i, \downarrow} a_{i, \uparrow} = \Delta_{0} + \delta\left(a_{i, \downarrow} a_{i, \uparrow}\right), \nonumber \\
&& b_{j, \sigma'} a_{i, \sigma} = \Delta_{\sigma', \sigma} \left({\bf r}_j, {\bf r}_i\right) + \delta\left(b_{j, \sigma'} a_{i, \sigma}\right), \nonumber
\end{eqnarray}
such that
\begin{eqnarray}
&& H_{{\rm eff}, U, V} = 2 \tilde{U} \sum_i \left|\Delta_0\right|^2 + \tilde{V} \sum_{\left<i,j\right>} \left|\Delta_{\sigma', \sigma} \left({\bf r}_j, {\bf r}_i \right)\right|^2 \nonumber \\
&&- \tilde{U} \sum_{i} \left[\Delta^{\dagger}_0 \left(a_{i, \downarrow} a_{i, \uparrow} + b_{i, \downarrow} b_{i, \uparrow} \right) + h.c. \right] \nonumber \\
&& - \tilde{V} \sum_{\left< i,j \right>} \left(\Delta^{\dagger}_{\sigma, \sigma'} \left({\bf r}_i, {\bf r}_j \right) b_{j, \sigma'} a_{i, \sigma} + h.c. \right), \label{eqn theeffectiveuandvhamiltonianafterdecoupling}
\end{eqnarray}
where the term quadratic in fluctuations $\mathcal{O}\left(\delta^2 \right)$ is neglected.

\subsection{Hamiltonian in Dirac-Nambu Representation}

We now transform the full Hamiltonian in Eq.~(\ref{eqn fullhamiltonianstatedinbegining}) with the $U$ and $V$ terms replaced by Eq.~(\ref{eqn theeffectiveuandvhamiltonianafterdecoupling}) to reciprocal space, expand around the Dirac points $\pm {\bf K}$, and use Dirac-Nambu representation to write the total Hamilltonain as
\begin{equation}
H = E_0 + \frac{1}{2} \sum_{{\bf q}} \Psi^\dagger M \Psi,
\end{equation}
where $E_0$ is the energy of the condensate and the $16$-component Dirac-Nambu spinors $\Psi^{\dagger} = \left(\Psi^{\dagger}_p, \Psi^{\dagger}_h \right)$, with $\Psi^{\dagger}_p = \left(\Psi^{\dagger}_{p \uparrow}, \Psi^{\dagger}_{p \downarrow}\right)$ and $\Psi^{\dagger}_h = \left(\Psi^{\dagger}_{h \downarrow}, -\Psi^{\dagger}_{h \uparrow}\right)$ are \cite{roy}
\begin{eqnarray}
\Psi^{\dagger}_{p, \sigma} ({\bf q}) &=& \begin{pmatrix}
 a^{\dagger}_{{\bf K} + {\bf q}, \sigma} & b^{\dagger}_{{\bf K} + {\bf q}, \sigma} & a^{\dagger}_{-{\bf K} + {\bf q}, \sigma} & b^{\dagger}_{-{\bf K} + {\bf q}, \sigma}
 \end{pmatrix}, \nonumber \\
\Psi^{\dagger}_{h, \sigma} ({\bf q}) &=& \begin{pmatrix}
 b_{{\bf K} - {\bf q}, \sigma} & a_{{\bf K} - {\bf q}, \sigma} & b_{-{\bf K} - {\bf q}, \sigma} & a_{-{\bf K} - {\bf q}, \sigma}
 \end{pmatrix}. \nonumber
\end{eqnarray}
The matrix $M$ is given in terms of the $16\times16$ matrices
\begin{equation}
\Gamma_{ijk} = \tau_i \otimes \sigma_j \otimes \gamma_k, \nonumber
\end{equation}
where $\tau_i$ and $\sigma_j$ are Pauli matrices acting in the particle-hole and spin space, respectively, and $\gamma_k$ are $4\times4$ matrices acting in the sublattice-valley space defined as $\gamma_0 = \sigma_0 \otimes \sigma_3$, $\gamma_1 = \sigma_3 \otimes \sigma_2$, $\gamma_2 = \sigma_0 \otimes \sigma_1$, $\gamma_3 = \sigma_1 \otimes \sigma_2$ and $\gamma_5 = \sigma_2 \otimes \sigma_2$.

\subsubsection{Dirac Hamiltonian}

The hopping term in Eq.~(\ref{eqn hubbardhamiltonianinrealspace}), can be written as $H_D = (1/2) \sum_{{\bf q}} \Psi^\dagger M_D \Psi$ with
\begin{equation}
M_D = v_F \, \tau_0 \otimes \sigma_0 \otimes i \gamma_0 \gamma_i q_i, \label{eqn dirachamiltoniancompactified}
\end{equation}
where $q_i = (q_y, - q_x)$ and $v_F = 3at/2$ is the Fermi velocity.

The Hamiltonian with the chemical potential in Eq.~(\ref{eqn reciprocalspacehamiltonianforthechemicalpotential}) is $H_\mu = (1/2) \sum_{{\bf q}} \Psi^\dagger M_\mu \Psi$ with
\begin{equation}
M_\mu = - \mu \, \tau_3 \otimes \sigma_0 \otimes \mathbb{I}. \label{eqn chemicalpotentialhamiltoniancompactified}
\end{equation}

\subsubsection{On-Site Pairing Hamiltonian}

Only considering the terms in Eq.~(\ref{eqn theeffectiveuandvhamiltonianafterdecoupling}) that include the on-site order parameter $\Delta_0$ leads to $H_{\Delta_0} = 4N \tilde{U} |\Delta_0|^2 + (1/2) \sum_{{\bf q}} \Psi^\dagger M_{\Delta_0} \Psi$ with
\begin{equation}
M_{\Delta_0} = - \tilde{U} \left[{\rm Re}\left(\Delta_0\right) \tau_1 - {\rm Im}\left(\Delta_0\right) \tau_2 \right] \otimes \sigma_0 \otimes i \gamma_0 \gamma_3. \label{eqn onsitecouplinghamiltoniancompactified}
\end{equation}

\subsubsection{Nearest-Neighbor Pairing Hamiltonian}

For the nearest-neighbor coupling, we use the Kekule ansatz\cite{roy}
\begin{equation}
\Delta_{\sigma, \sigma} \left({\bf r}_i, {\bf r}_j\right) = \Delta_{\sigma} \, {\rm cos} \left({\bf K} \cdot \left({\bf r}_i + {\bf r}_j\right)\right), \label{eqn kekuleansatzpartone}
\end{equation}
\begin{equation}
\frac{1}{2}\left(\Delta_{\downarrow, \uparrow} \left({\bf r}_i, {\bf r}_j\right) +\Delta_{\uparrow, \downarrow} \left({\bf r}_i, {\bf r}_j\right) \right) = \Delta \, {\rm cos} \left({\bf K} \cdot \left({\bf r}_i + {\bf r}_j\right) \right), \label{eqn kekuleansatzparttwo}
\end{equation}
\begin{equation}
\frac{1}{2} \left(\Delta_{\downarrow, \uparrow} \left({\bf r}_i, {\bf r}_j\right) - \Delta_{\uparrow, \downarrow} \left({\bf r}_i, {\bf r}_j\right) \right) = \Delta', \label{eqn kekuleansatzpartthree}
\end{equation}
where Eqs.~(\ref{eqn kekuleansatzpartone}) and (\ref{eqn kekuleansatzparttwo}) represent a spin-triplet Kekule order, while Eq.~(\ref{eqn kekuleansatzpartthree}) represents a spin singlet, the so-called hidden order. \cite{uchoa} In its full generality, the Kekule ansatz contains a phase. However, in the Dirac approximation we use here, i.e. only including electrons close to the Dirac points, the results are independent of this phase, and we have set it to zero. This degeneracy is, however, (weakly) broken when the lattice is reintroduced in the problem.\cite{roy}

Inserting this ansatz into Eq.~(\ref{eqn theeffectiveuandvhamiltonianafterdecoupling}) leads to the following mean-field Hamiltonian for the Kekule order
\begin{equation}\label{Ham-Kekule}
H_{\rm Kekule} = 6 N \tilde{V} m^2 + \frac{1}{2} \sum_{{\bf q}} \Psi^\dagger M_m \Psi,
\end{equation}
with
\begin{eqnarray}
&& M_m = - \tilde{V} \left[\left(X \tau_1 - Y \tau_2 \right) \otimes \sigma_3 + \left(I_- \tau_2 - R_- \tau_1 \right) \otimes \sigma_1 \right. \nonumber \\
&& \left. + \left(I_+ \tau_1 + R_+ \tau_2 \right) \otimes \sigma_2 \right] \otimes \gamma_0. \label{eqn kekuleorderhamiltoniancompactified}
\end{eqnarray}
 Here, $\Delta = X + i Y$, $R_{\pm} = \frac{1}{2} \left[{\rm Re}\left(\Delta_{\uparrow}\right) \pm {\rm Re}\left(\Delta_{\downarrow}\right) \right]$, $I_{\pm} = \frac{1}{2} \left[{\rm Im}\left(\Delta_{\uparrow}\right) \pm {\rm Im}\left(\Delta_{\downarrow}\right) \right]$ and
\begin{equation}
m^2 = X^2 + Y^2 +R_+^2 + I_+^2 + R_-^2 + I_-^2.
\end{equation}

For the hidden order, we find
\begin{equation}\label{Ham-hidden}
H_{\Delta'} = 12 N \tilde{V} |\Delta'|^2 + \frac{1}{2} \sum_{{\bf q}} \Psi^\dagger M_{\Delta'} \Psi,
\end{equation}
with
\begin{equation}
M_{\Delta'} = \frac{2i}{t} \tilde{V} \left[{\rm Re}\left(\Delta'\right) \tau_2 + {\rm Im} \left(\Delta'\right) \tau_1 \right] \otimes \sigma_1 \otimes i \gamma_0 \gamma_3 M_D. \label{eqn hiddenorderhamiltoniancompactified}
\end{equation}
The proportionality to the Dirac Hamiltonian indicates that instead of opening a superconducting gap, the hidden order renormalizes the Fermi velocity $v_F$.

\section{Competition between Kekule and hidden orders} \label{section 4}

 From now on, we only consider the Kekule and hidden order by setting the on-site order parameter to zero, i.e. $\Delta_0 = 0$. This case is more interesting to study because when all three superconducting orders are included, the $s$-wave order parameter is preferred, as is discussed in App.~\ref{appendixsection swavescreviewed}. We start by first considering the Kekule and hidden order parameters separately (excluding the possibility of their coexictence) in the gap equations at both finite and zero temperature. Finally, we solve  the self-consistent mean-field gap equations analytically at zero temperature and half filling.

\subsection{Gap equations}

First, we derive the thermodynamical potential for our system, which is followed by the computation of the gap equations and critical couplings.

\begin{figure}[h!]
\centering
\subfigure{
\includegraphics[width=0.45\textwidth]{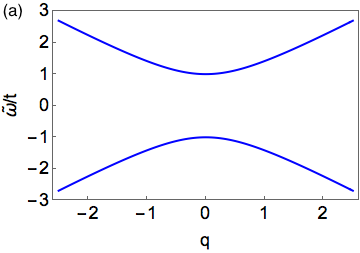}}~

\subfigure{
\includegraphics[width=0.45\textwidth]{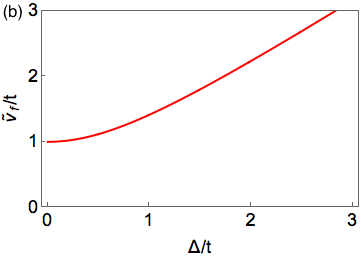}}
\caption{(Color online) The dispersion relation in Eq.~(\ref{eqn dispersionrelationforkekuleandhidden}) is plotted for $\mu/t = \Delta'/t = 0$, $\tilde{V}/t = v_{F}/t = 1$ and $m = 1$ (panel (a)), showing that the Kekule order $m$ opens a gap. In panel (b), the renormalization of the Fermi velocity by the hidden order parameter, given by Eq.~(\ref{vF-renormalization}), is shown. \label{fig dispersionrelation}}
\end{figure}

\begin{figure}[b!]
\centering
\subfigure{
\includegraphics[width=0.45\textwidth]{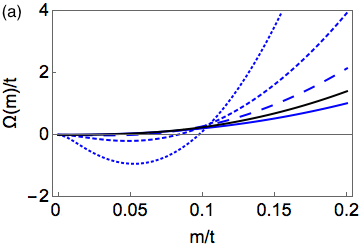}}~

\subfigure{
\includegraphics[width=0.45\textwidth]{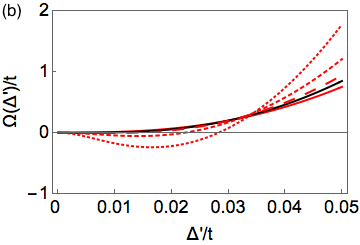}}
\caption{(Color online) Evolution of the thermodynamical potential in Eq.~(\ref{eqn thermodynamicalpotential}) for the Kekule (a) and hidden order (b) with $v_F/t = \Lambda = N = 1$, $k_B T/t = 0.1$, $\mu/t=0$. (a) Kekule order: $\tilde{V}/t = 2 \pi$ (solid), $\tilde{V}/t = 3 \pi$ (black), $\tilde{V}/t = 5 \pi$ (large dashed), $\tilde{V}/t = 10 \pi$ (small dashed), and $\tilde{V}/t = 25 \pi$ (dotted). (b) Hidden order: $\tilde{V}/t = 15 \pi$ (solid), $\tilde{V}/t = 18 \pi$ ( black), $\tilde{V}/t = 21 \pi$ (large dashed), $\tilde{V}/t = 30 \pi$ (small dashed), and $\tilde{V}/t = 50 \pi$ (dotted).} \label{fig freeenergyplotsshowingsecondorderphasetransition}
\end{figure}

\subsubsection{Thermodynamical Potential}

The thermodynamical potential $\Omega$ is obtained using the partition function
\begin{equation}
Z = {\rm e}^{- \beta \Omega} = {\rm Tr}\left({\rm e}^{- \beta H_{\rm tot}}\right),
\end{equation}
where $H_{\rm tot}$ is the sum of the Hamiltonians in Eqs.~(\ref{eqn dirachamiltoniancompactified})-(\ref{eqn onsitecouplinghamiltoniancompactified}), (\ref{Ham-Kekule}) and (\ref{Ham-hidden}). Performing the trace yields
\begin{equation}
\Omega = E_0 - \frac{1}{\beta} \sum_{{\bf q}; s, s'= \pm} {\rm ln} \left(1 + {\rm e}^{-\beta \tilde{\omega}_{s, s'}} \right), \label{eqn thermodynamicalpotential}
\end{equation}
where
\begin{equation}
E_0 = 6N \tilde{V} \left(m^2 + 2 \left|\Delta'\right|^2 \right),
\end{equation}
and $\tilde{\omega}_{s, s'} = s \tilde{\omega}_{s'}$ is obtained by diagonalizing $M_D + M_\mu + M_m + M_{\Delta'}$, with
\begin{equation}
\tilde{\omega}_{s'} = \sqrt{\left(v_F |q| + s' \mu\right)^2 + \tilde{V}^2 m^2 + \frac{v_F^2}{t^2} |q|^2 \tilde{V}^2 \left|\Delta'\right|^2},
\label{eqn dispersionrelationforkekuleandhidden}
\end{equation}
where $s=\pm1$ and $s'=\pm1$ correspond to the spin and particle-hole degree of freedom, respectively. From this dispersion, we can see that the Kekule order $m$ acts as a mass term for the Dirac fermions, and opens a superconducting gap as shown in Fig.~\ref{fig dispersionrelation} (a). The hidden order, on the other hand, for $\mu = 0$, renormalizes Fermi velocity according to
\begin{equation}\label{vF-renormalization}
\tilde{v}_F = t \sqrt{1 + \tilde{V}^2 |\Delta'|^2/t^2},
\end{equation}
as is displayed in Fig.~\ref{fig dispersionrelation} (b).
In Fig.~\ref{fig freeenergyplotsshowingsecondorderphasetransition}, the evolution of the thermodynamical potential with respect to different couplings is shown for both the Kekule and hidden order. We clearly see a second-order phase transition, with the thermodynamic potential at the critical coupling for the quantum phase transition shown in black.

\subsubsection{Finite-Temperature Gap Equations}

Minimizing Eq.~(\ref{eqn thermodynamicalpotential}) with respect to the Kekule order parameter leads to the finite-temperature gap equation
\begin{equation}
1 = \frac{\tilde{V}}{3} \sum_{s = \pm} \int \frac{{\rm d}{\bf q}}{(2 \pi)^2} \frac{1}{\tilde{\omega}_s} {\rm tanh} \left(\frac{\beta \tilde{\omega}_s}{2} \right), \label{eqn finitetemperaturegapequationkekule}
\end{equation}
where $\tilde{\omega}_{s}$ is given by Eq.~(\ref{eqn dispersionrelationforkekuleandhidden}). The equation determining the hidden order parameter at finite-temperature, which we loosely also call gap equation, is analogously obtained
\begin{equation}
1 = \frac{\tilde{V}}{6} \frac{v_F^2}{t^2} \sum_{s = \pm} \int \frac{{\rm d}{\bf q}}{(2 \pi)^2} \frac{|q|^2}{\tilde{\omega}_s} {\rm tanh} \left(\frac{\beta \tilde{\omega}_s}{2} \right). \label{eqn finitetemperaturegapequationhidden}
\end{equation}

\begin{figure}[b]
\centering
\includegraphics[width=0.4\textwidth]{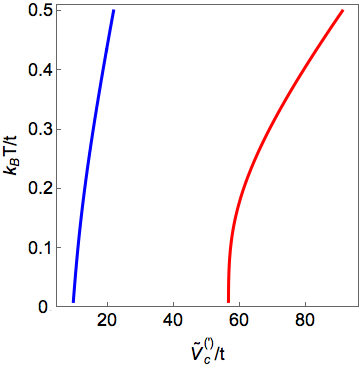}
\caption{(Color online.) Behavior of the critical interaction for the Kekule (blue) and hidden order (red) with increasing temperature with $v_F/t = \Lambda = 1$, $\mu/t = 0$, $m/t = \Delta'/t = 0$.}
\label{fig criticalcouplingbehaviorwithtemperature}
\end{figure}

\subsubsection{Critical Coupling}

At half-filling ($\mu = 0$) the Fermi energy is pinned at the Dirac points, and the density of states vanishes implying that there is a critical interaction at which the superconducting order sets in.
For the Kekule order, we obtain the critical coupling
\begin{equation}
\tilde{V}_c (T) = 3 \pi v_F \left[\Lambda - \frac{2 {\rm ln}(2)}{\beta v_F} \right]^{-1},
\end{equation}
such that the zero-temperature critical interaction  $\tilde{V}_c(0) \equiv \tilde{V}_c = 3 \pi v_F /\Lambda$. For the hidden order, we find
\begin{equation}\label{Vc-hidden}
 \tilde{V}'_c (T)= \frac{18 \pi t^2}{v_F} \left[\Lambda^3 - \frac{9}{\beta^3 v_F^3} \zeta(3) \right]^{-1},
\end{equation}
yielding for the zero-temperature critical interaction $\tilde{V}'_c(0) \equiv \tilde{V}'_c = 18 \pi t^2/(v_F \Lambda^3)$, with the corresponding integral for the hidden order solved in App.~\ref{appendixsection criticalinteractionhiddenorder}. Here, $\Lambda$ is the high-energy cutoff, up to which the continuum Dirac theory is valid, and which scales with the band-width of the order of the nearest-neighbor hopping.

Taking $v_F/t=\Lambda = 1$ and $T=0$, we see that the critical coupling for the Kekule order is smaller than the one for the hidden order. This implies that the system first enters the Kekule superconducting order and this state, therefore, dominates over the hidden order. In Fig.~\ref{fig criticalcouplingbehaviorwithtemperature}, we show how the critical interaction behaves with increasing temperature. We see that the critical interaction for the Kekule order remains smaller than that for the hidden order for any temperature. Furthermore, the function $\tilde{V_c}(T)$ ($\tilde{V'_c}(T)$) determines finite-temperature crossover from the quantum-critical semimetal to the Kekule (the hidden) superconducting state. \cite{sachdev-book}  Finally, we  observe that the critical interaction for both orders increases with temperature, since thermal fluctuations are expected to be detrimental for an ordered phase, meaning that the critical coupling should increase.

\subsection{Zero Temperature}

The zero-temperature gap equations for the Kekule and hidden order are obtained from Eqs.~(\ref{eqn finitetemperaturegapequationkekule}) and (\ref{eqn finitetemperaturegapequationhidden}), respectively, by setting $T=0$ such that
\begin{eqnarray}
1 &=& \frac{\tilde{V}}{3} \sum_{s = \pm} \int \frac{{\rm d}{\bf q}}{(2 \pi)^2} \frac{1}{\tilde{\omega}_s}, \label{eqn zerotemperaturegapequationkekuleorder} \\
1 &=& \frac{\tilde{V}}{6} \frac{v_F^2}{t^2} \sum_{s = \pm} \int \frac{{\rm d}{\bf q}}{(2 \pi)^2} \frac{|q|^2}{\tilde{\omega}_s}. \label{eqn zerotemperaturegapequationhiddenorder}
\end{eqnarray}
Solving these integrals for weak and strong  couplings, we can derive the zero-temperature gaps in  both these limits.

\subsubsection{Kekule Order}

To find the zero-temperature gap for the Kekule order, we set $\Delta' = 0$ in Eq.~(\ref{eqn zerotemperaturegapequationkekuleorder}), such that it simplifies to
\begin{eqnarray}
1 &=& \frac{\tilde{V}}{6 \pi v_F^2} \left[2 \left(v_F \Lambda - \sqrt{\mu^2 + \tilde{V}^2 m^2 (0)} \right) \right. \nonumber \\
&& \left. + \mu \, {\rm ln} \left(\frac{\mu + \sqrt{\mu^2 + \tilde{V}^2 m^2(0)}}{- \mu + \sqrt{\mu^2 + \tilde{V}^2 m^2(0)}} \right) \right]. \label{eqn zerotemperaturegapequationkekulerewritten}
\end{eqnarray}
At zero chemical potential, we find for the Kekule gap
\begin{equation}
m\left(0, \mu = 0 \right) = \frac{3 \pi v_F}{\tilde{V}_c \tilde{V}} \left(1 - \frac{\tilde{V}_c}{\tilde{V}}\right).
\end{equation}

Next, we solve the zero-temperature gap equation at finite chemical potential. Analytical solutions can only be found in the strong- and weak-coupling limit, $\tilde{V}>\tilde{V}_c$ with $m(0)/ \mu \gg 1$ and $\tilde{V}<\tilde{V}_c$ with $m(0)/ \mu \ll 1$, respectively. Note that in the strong-coupling limit $\mu \ll 1$, so that the Fermi level is in the vicinity of the Dirac points. Therefore, this limit is governed by the quantum-critical point and we expect the resulting zero-temperature gap to exhibit power-law behavior. In the weak-coupling limit, on the other hand, the zero-temperature gap is expected to have a BCS-like form, since the system is away from the quantum-critical regime with a finite density of states at the Fermi level. Applying these limits to Eq.~(\ref{eqn zerotemperaturegapequationkekulerewritten}) yields
\begin{widetext}
\begin{equation}
m \left(0, \mu \right) \rightarrow \begin{cases}
  \frac{m\left(0, \mu = 0\right)}{2} \left[1 + \sqrt{1 + \frac{4 \mu^2}{\tilde{V}_0^2 \, m\left(0, \mu = 0 \right)^2}} \right],      & \quad \tilde{V} > \tilde{V}_c, \, m(0)/\mu \gg 1,\\
    \frac{2 \mu}{\tilde{V}}\, \exp{[{\frac{\tilde{V}}{\mu} m\left(0, \mu = 0\right) -1}]}, & \quad \tilde{V} < \tilde{V}_c, \, m(0)/\mu \ll 1.\\
  \end{cases} \nonumber
\end{equation}
As expected, the superconducting gap shows quantum-critical power-law behavior at the strong-coupling, and is BCS-like in the weak-coupling limit.

\begin{figure}[t!]
\centering
\subfigure{
\includegraphics[width=0.45\textwidth]{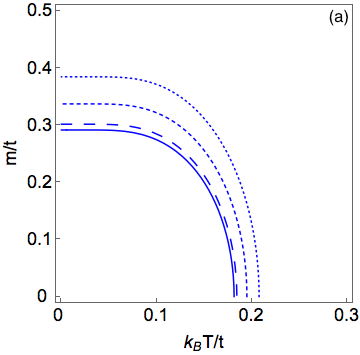} \label{fig solutiontothefinitetemperaturegapequatiosan}}~
\subfigure{
\includegraphics[width=0.45\textwidth]{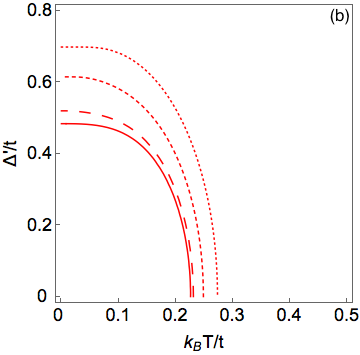} \label{fig solutiontothefinitetemperaturegapequatiosbn}}
\caption{(Color online) Solutions of the finite-temperature gap equation for the Kekule (a) and hidden order (b) with $v_F/t = \Lambda = 1$. In panel (a) $\Delta'/t = 0$ and $\tilde{V}/t = 4 \pi$, while in panel (b) $m/t = 0$ and $\tilde{V}/t = 20 \pi$. In the plots we use the following values of the chemical potential: $\mu/t =0$ (solid), $\mu/t = 0.1$ (large dashed), $\mu/t = 0.25$ (small dashed), and $\mu/t = 0.5$ (dotted).}
\label{fig solutiontothefinitetemperaturegapequatios}
\end{figure}

\subsubsection{Hidden Order}

We solve Eq.~(\ref{eqn zerotemperaturegapequationhiddenorder}) for $m = 0$ and obtain
\begin{equation}
1 = \frac{\tilde{V}}{6} \frac{v_F}{t^2} \frac{1}{2\pi} \left\{2 \Lambda \left[\frac{\Lambda^2}{3 \sqrt{1 + \alpha^2}} + \frac{\mu^2}{v_F^2} \left(\frac{3 - \alpha^2}{\left(1 + \alpha^2 \right)^{5/2}} \right) \right] + \frac{\mu^3}{v_F^3} \left[\frac{4 \alpha^2 - 11}{9 \left(1 + \alpha^2\right)^3} + {\rm ln} \left(\frac{\sqrt{1 + \alpha^2}+1}{\alpha} \right) \left(\frac{2 - 3 \alpha^2}{\left(1 + \alpha^2\right)^{7/2}} \right) \right] \right\}, \label{eqn resultofsolvingthezerotemperaturegapforthehiddenorder}
\end{equation}
\end{widetext}
where $\alpha \equiv \tilde{V} |\Delta'(0)|/t$. Details of the calculation are presented in App.~\ref{appendixsection solvingthezerotemperaturehiddengapequation}. For the hidden order parameter at zero chemical potential, we find
\begin{equation}
\left|\Delta'\left(0, \mu = 0\right)\right| = \frac{t}{\tilde{V}'_c} \sqrt{1 - \left(\frac{\tilde{V}'_c}{\tilde{V}}\right)^2},
\end{equation}
which scales with $t$ suggesting that this order parameter in fact does not open a gap, but renormalizes the Fermi velocity.

At finite chemical potential, we find a solution in the strong- and weak-coupling limit, where $|\Delta'(0)|/t \gg 1$ and $|\Delta'(0)|/t \ll 1$, respectively, such that
\begin{equation}
\left|\Delta'\right| \left(0, \mu \right) \rightarrow \begin{cases}
    \frac{t}{\tilde{V}'_c}, & \tilde{V} > \tilde{V}'_c, \, |\Delta'(0)|/t \gg 1,\\
     \frac{2 t}{\tilde{V}} \, \exp{\left[{ F\left(\tilde{V}, \tilde{V}'_c \right)}\right]}, &  \tilde{V} < \tilde{V}'_c, \, |\Delta'(0)|/t \ll 1,\\
  \end{cases} \nonumber
\end{equation}
with $F\left(\tilde{V}, \tilde{V}'_c \right) = \frac{6 \pi t^2 v_F^2}{\tilde{V}'_c \mu^3} \left(1 - \frac{\tilde{V}'_c}{\tilde{V}} \right) - \frac{11}{18}$.

\subsection{Critical Temperature}

Lastly, we determine the crossover (loosely called ``critical" hereafter) temperatures in the strong- and weak-coupling limit from the corresponding finite-temperature gap equations (\ref{eqn finitetemperaturegapequationkekule}) and (\ref{eqn finitetemperaturegapequationhidden}), respectively, by requiring that $m(T_c)=0$ ($\Delta'(T'_c) = 0$) at the transition into the Kekule (hidden) order.

\subsubsection{Kekule order}

The solution of Eq.~(\ref{eqn finitetemperaturegapequationkekule}) at the critical temperature for $\Delta' = 0$ is shown in Fig.~\ref{fig solutiontothefinitetemperaturegapequatiosan}. The critical temperature increases with increasing chemical potential, as expected from the fact that the density of states scales linearly with the chemical potential. An explicit expression for the critical temperature can be derived in the strong- and weak-coupling limit
\[ T_c \rightarrow \left\{
  \begin{array}{l l}
    \frac{1}{2 \, {\rm ln}(2) k_B} \left[\frac{\tilde{V}^2 m(0, \mu)^2}{\mu + \tilde{V}^2 m(0, \mu)} + \mu \right], & \quad \tilde{V} > \tilde{V}_c, \, \beta_c \mu \ll 1,\\
    \frac{e^\gamma}{k_B \pi} \tilde{V} m(0, \mu), & \quad \tilde{V} < \tilde{V}_c, \, \beta_c \mu \gg 1,
  \end{array} \right.\]
which again shows power-law behavior in the strong-coupling limit and BCS-like behavior in the weak-coupling limit.

\begin{figure}[b!]
\centering
\includegraphics[width=0.4\textwidth]{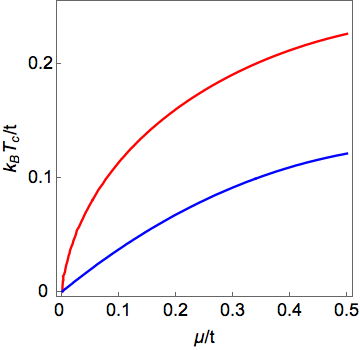}
\caption{(Color online) Behavior of the critical temperature for the Kekule (blue) and hidden order (red) with increasing chemical potential with $v_F/t = \Lambda = 1$. For the Kekule order (blue), we chose $\tilde{V}/t = 3 \pi$, and for the hidden order (red) $\tilde{V}/t = 18 \pi$.}
\label{fig criticaltemperatureversuschemicalpotential}
\end{figure}

\subsubsection{Hidden Order}

We solve the finite-temperature gap equation for the hidden order in Eq.~(\ref{eqn finitetemperaturegapequationhidden}) in a similar fashion by setting $m=0$ and requiring $\Delta'(T'_c) = 0$ at the transition. The solution is shown in Fig.~\ref{fig solutiontothefinitetemperaturegapequatiosbn} with features similar to Fig.~\ref{fig solutiontothefinitetemperaturegapequatiosan}. The integral is solved in a similar fashion (outlined in App.~\ref{appendixsection solvingthezerotemperaturehiddengapequation}) and the explicit expression for the critical temperature reads
\begin{widetext}
\begin{equation}
T'_c \rightarrow \begin{cases}
    \frac{1}{k_B} \left[\frac{2 \pi t^2 v_F^2}{\zeta(3) \tilde{V}} \left(\frac{\tilde{V} \left|\Delta'(0, \mu)\right|}{t} - 1 \right) \right]^{1/3}, & \, \tilde{V} > \tilde{V}'_c, \, \beta_c \mu \ll 1,\\
     \frac{\mu {\rm e}^{\gamma - \frac{11}{9}}}{k_B \pi} \frac{\tilde{V} \left|\Delta'(0, \mu)\right|}{t}, & \, \tilde{V} < \tilde{V}'_c, \, \beta_c \mu \gg 1.\\
  \end{cases} \nonumber
\end{equation}

Fig.~\ref{fig criticaltemperatureversuschemicalpotential} shows that the critical temperature for both the Kekule and hidden order parameters increases with chemical potential. Moreover, in Fig.~\ref{fig logcriticaltemperatureversuscoupling} the behavior of the critical temperature for both the Kekule and hidden order as a function of the coupling for different values of the chemical potential is displayed. We observe that the critical coupling decreases with increasing chemical potential. These features are expected on physical grounds, since the density of states linearly increases with energy as one moves away from the Dirac points.
\begin{figure}[t!]
\centering
\subfigure{
\includegraphics[width=0.45\textwidth]{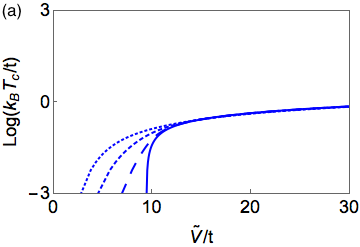} \label{fig solutiontothefinitetemperaturegapequatiosa}}~
\subfigure{
\includegraphics[width=0.45\textwidth]{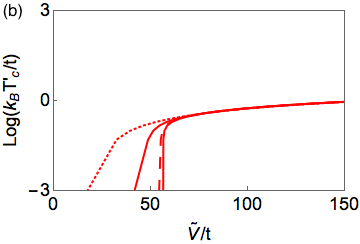} \label{fig solutiontothefinitetemperaturegapequatiosb}}
\caption{(Color online) Solutions of the finite-temperature gap equation for the Kekule (a) and hidden order (b) with $v_F/t = \Lambda = 1$ for different values of the chemical potential: $\mu/t = 0$ (solid), $\mu/t = 0.1$ (large dashed), $\mu/t = 0.25$ (small dashed) and $\mu/t = 0.5$ (dotted). $m(T_c)$ = 0 and $\Delta'/t = 0$ for (a), and $\Delta'(T'_c) = 0$ and $m/t = 0$ for (b).}
\label{fig logcriticaltemperatureversuscoupling}
\end{figure}
\end{widetext}

\begin{figure}[b!]
\centering
\includegraphics[width=0.4\textwidth]{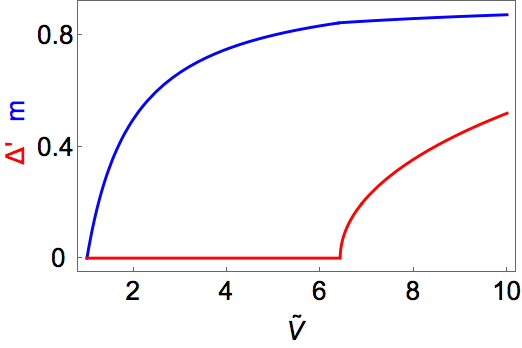}
\caption{(Color online) Behavior of the Kekule (blue) and hidden (red) order parameters with interaction strength ${\tilde V}$, given by Eq.\ (\ref{Kekule-competition}) and (\ref{hidden-competition}), respectively. We use the critical coupling for the Kekule and hidden order in the absence of the competition $\tilde{V_c} = 1$ and $\tilde{V'_c} = 5$, respectively.}
\label{fig evolutionoftwoopsasfctofinteractionstrength}
\end{figure}

\subsection{Self-consistent gap equations at zero temperature and at half filling}

We now consider the competition of the Kekule and the hidden superconducting orders within the framework of the self-consistent mean-field gap equations (\ref{eqn finitetemperaturegapequationkekule}) and (\ref{eqn finitetemperaturegapequationhidden}) at zero temperature and zero chemical potential, which, after integrating over the angle, are rewritten as
\begin{align}
&1=\frac{\tilde{V}}{3\pi}\int_0^{\Lambda} dq\frac{q}{\sqrt{q^2+m^2+q^2|\Delta'|^2}},\\
&1=\frac{\tilde{V}}{6\pi}\int_0^{\Lambda} dq\frac{q^3}{\sqrt{q^2+m^2+q^2|\Delta'|^2}}.
\end{align}
Here, we have conveniently redefined ${\tilde V}^2 m^2 \rightarrow m^2$ and $v_F^2{\tilde V}^2/t^2 |\Delta'|^2\rightarrow |\Delta'|^2$.
By rescaling the momentum, $q\rightarrow q(1+|\Delta'|^2)^{1/2}$, and performing the integration, we obtain
\begin{align}
&1=\frac{\tilde{V}}{\tilde{V_c}}(1+|\Delta'|^2)^{-1}(1-m),\label{eq:kekulecomp}\\
&1=\frac{\tilde{V}}{\tilde{V'_c}}(1+|\Delta'|^2)^{-2}(1-2m^2+2m^3),
\label{eq:hiddencomp}
\end{align}
where we redefined $m/\Lambda \rightarrow m$. By inserting Eq.~(\ref{eq:kekulecomp}) into Eq.\ (\ref{eq:hiddencomp}), and solving for the hidden order parameter we obtain six solutions of which only one is physically relevant. This solution has the form
\begin{widetext}

\begin{equation}\label{hidden-competition}
\Delta'(\tilde{V},\tilde{V_c},\tilde{V'_c})=\sqrt{\frac{\tilde{V} \left[4 \tilde{V_c}{}^2 \left(G^{1/3}-2
   \tilde{V} \tilde{V'_c}\right)-G^{1/3} \tilde{V}
   \tilde{V'_c}+\tilde{V}^2 \tilde{V'_c}{}^2+4
   \tilde{V_c}{}^4+G^{2/3}\right]}{6\tilde{V_c}{}^3 {G}^{1/3}}-1},
\end{equation}
and yields the following Kekule gap
\begin{equation}\label{Kekule-competition}
m(\tilde{V},\tilde{V_c},\tilde{V'_c})=\frac{2 \tilde{V_c}{}^2 \left(4 \tilde{V}
   \tilde{V'_c}+{G}^{1/3}\right)+{G}^{1/3} \tilde{V}
   \tilde{V'_c}-\tilde{V}^2 \tilde{V'_c}{}^2-4
   \tilde{V_c}{}^4-G^{2/3}}{6  \tilde{V_c}{}^2 {G}^{1/3}},
\end{equation}

\noindent with the function $G\equiv G(\tilde{V},\tilde{V_c},\tilde{V'_c})$ defined as
\begin{equation}
G(\tilde{V},\tilde{V_c},\tilde{V'_c})=-30 \tilde{V} \tilde{V_c}{}^4 \tilde{V'_c}+12 \tilde{V}^2 \tilde{V_c}{}^2
   \tilde{V'_c}{}^2-\tilde{V}^3 \tilde{V'_c}{}^3+46
   \tilde{V_c}{}^6+6
   \tilde{V_c}{}^3 \sqrt{-66 \tilde{V} \tilde{V_c}{}^4
   \tilde{V'_c}+33 \tilde{V}^2 \tilde{V_c}{}^2
   \tilde{V'_c}{}^2-3 \tilde{V}^3 \tilde{V'_c}{}^3+57
   \tilde{V_c}{}^6}.
   \end{equation}
\end{widetext}

\begin{figure}[b!]
\centering
\includegraphics[width=0.4\textwidth]{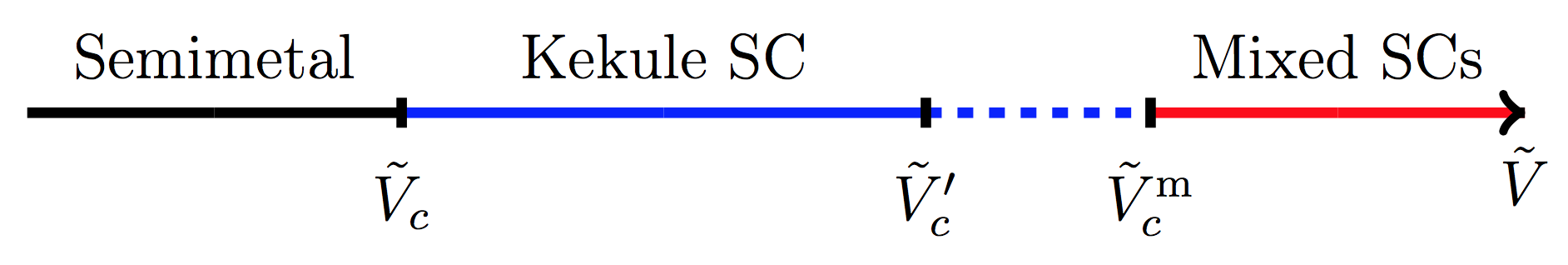}
\caption{(Color online.) Phase diagram of the system as a function of the nearest-neighbor attraction $\tilde{V}$. As this coupling increases, at a critical value $\tilde{V_c}$ the system first enters the Kekule superconducting state (blue region). In the region labeled by dashed blue lines, the Kekule order is favored over the hidden order. The latter would in the absence of the Kekule order set in for $\tilde{V}>\tilde{V'_c}$, but eventually coexists with the Kekule superconductor above the critical interaction $\tilde{V_c}^{\rm m}>\tilde{V'_c}$ (red solid line). }
\label{fig phasediagram}
\end{figure}

\begin{figure}[tb]
\centering
\includegraphics[width=0.4\textwidth]{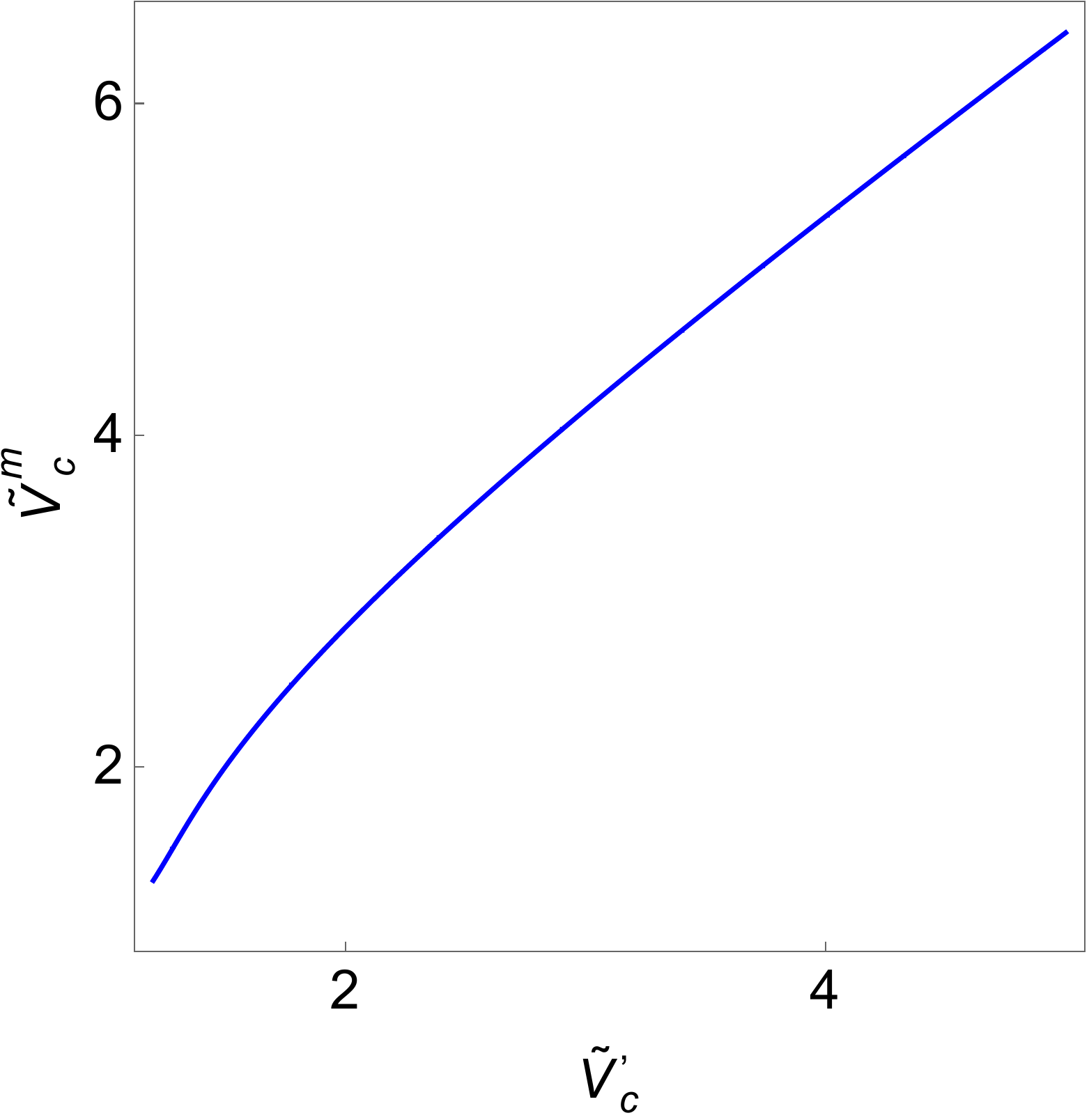}
\caption{(Color online.) Dependence of the critical coupling for the onset of the mixed phase $\tilde{V_c}^{\rm m}$ on the critical coupling for the hidden order in absence of the competition, $\tilde{V'_c}$. The critical coupling for the Kekule order $\tilde{V}_c/\Lambda = 1$.}
\label{fig mixedcriticalvsbarecritical}
\end{figure}

As a result, above a critical value of the nearest-neighbor attraction $\tilde{V_c}^{\rm m}$, we obtain a phase in which the Kekule and the hidden superconducting orders {\it coexist}. In Fig.~\ref{fig evolutionoftwoopsasfctofinteractionstrength} we plot the Kekule gap and the hidden order parameter as a function of the nearest-neighbor attraction for fixed values of the critical couplings for the "bare" Kekule and the hidden orders, i.e, the critical couplings obtained without taking into account their competition. We observe that the value of the critical interaction $\tilde{V_c}^{\rm m}$ is greater than the bare value for the hidden superconducting state. Hence, in the region $\tilde{V'_c}<\tilde{V}<\tilde{V_c}^{\rm m}$ sketched in Fig.~\ref{fig phasediagram}, the Kekule superconductor is favored over the hidden order.
However, the latter eventually sets in, and the two orders coexist. The dependence of the critical coupling for the mixed phase on the bare critical coupling for the hidden order is shown in Fig.~\ref{fig mixedcriticalvsbarecritical}. We observe that $\tilde{V_c}^{\rm m}>\tilde{V}'_c$, expected based on the fact that the Kekule superconductor is gapped, while the hidden is gapless, which makes the former favorable over the latter. The preference of the Kekule order over the hidden order is so strong that for larger couplings, the system favors a mixed phase over a phase with a purely hidden superconducting order. Furthermore, the transition from the Kekule into the mixed phase is of the second order, which consists of the two second order transitions in the separate Kekule and hidden order channels. This feature may be an artifact of the mean-field approximation and when the fluctuations are included, this transition may turn out to be of the first order. However, this problem is beyond the scope of this work, and will be addressed in the future.

\section{Discussion and Conclusions} \label{section 5}

Here, we show explicit values for the effective Hubbard terms in Eqs.~(\ref{eqn effectivehubbarduinteraction}) and (\ref{eqn effectivehubbardvinteraction}) obtained by describing the coupling of the electrons with the LO lattice deformations using a continuum dielectric model. The semiconductor sheet is defined by its static $\varepsilon_{\rm in}(0)$ and optical (high frequency) $\varepsilon_{\rm in}(\infty)$ dielectric constants. This approximation is commonly used to describe polarons in ionic materials \cite{Pekar46,Frolich50} and LO-phonon coupling in semiconductor nanocrystals.\cite{Klein90,Delerue04,sun} The capping dielectric layer, if present, is described by $\varepsilon_{\rm out}(0)$ and $\varepsilon_{\rm out}(\infty)$. Details of the calculations are given in App.~\ref{appendix_couplings}.

\begin{table}
\caption{\label{table} Parameters (in meV) defining the effective interactions in the free-standing superlattices of CdSe or PbSe, and in the superlattice of CdSe capped with LiF. $\hbar \omega_E$ is the energy of the LO phonon which gives the strongest coupling to the electrons.}
\begin{ruledtabular}
\begin{tabular}{cccccccc}
System & $\hbar \omega_E$ & $U$ & $V$ & $V_0$ & $\tilde{V}_0$ & $\hbar \beta N \tilde{U}$ & $\hbar \beta N \tilde{V}$ \\ \hline
CdSe & 26 & 496 & 262 & 36 & 7 & -360 & -666 \\
PbSe & 17 & 290 & 208 & 32 & 12 & -6 & -343 \\
CdSe/LiF & 82 & 148 & 66 & 53 & 27 & 78 & 9
\end{tabular}
\end{ruledtabular}
\end{table}

Table~\ref{table} summarizes the results of the numerical calculations performed on the superlattices described in Ref.~\onlinecite{kalesaki}. We consider nanocrystals with a truncated cubic shape and a size of 4.3 nm assembled in a honeycomb lattice. The $\langle 111 \rangle$ axis of the atomic lattice is oriented perpendicularly to the plane of the honeycomb sheet. The Coulomb interactions are obtained using $\varepsilon_{\rm in}(0) = 10$ and $\varepsilon_{\rm in}(\infty) = 6$ for CdSe, and $\varepsilon_{\rm in}(0) = 280$ and $\varepsilon_{\rm in}(\infty) = 25.2$ for PbSe. When the superlattices are free-standing, the effective interactions $\tilde{U}$ and $\tilde{V}$ remain repulsive, even though their magnitude is strongly reduced due to the coupling to phonons. In the case of PbSe, the effect is stonger because it is characterized by higher dielectric constants than CdSe. In fact, the effective interactions always remain repulsive due to the leakage of a large part of the electric field into the vacuum surrounding the superlattice when an electron is placed in a nanocrystal, implying that the dielectric screening from the ionic polarization is unable to overturn the initial repulsive interaction. The opposite situation occurs when the CdSe superlattice is placed at $0.5$ nm from a semi-infinite LiF sample [$\varepsilon_{\rm out}(0) = 8.9$ and $\varepsilon_{\rm out}(\infty) = 1.9$]. In that case, the effective interaction becomes positive due to the electric field that strongly penetrates the external dielectric, and the main coupling comes from its polarization. The contribution coming from the polarization of CdSe can be neglected in a first approximation and $\hbar \omega_E$ can be identified with its value in LiF. However, even in this case the on-site attractive interaction remains dominant over the nearest-neighbor one (see Table I and notice that positive values of energy actually correspond to attractive interactions because of an overall minus sign in the Hamiltonian). Despite that, we theoretically investigated the more exotic superconducting orders that may arise when the nearest-neighbor attractive interaction dominates over the on-site one, which may also be relevant in the context of the ultracold atom systems. \cite{lih}

More specifically, we have investigated the problem of the competition between the Kekule and hidden superconducting orders in self-assembled artificial nanocrystals of graphene, at both finite temperature and finite chemical potential, within the mean-field theory for Dirac electrons, first by excluding the possibility of their coexistence. As a result of this analysis, we find that the Kekule order is preferred over the hidden order at both a finite temperature and a finite chemical potential. On the other hand, within the self-consistent mean-field approximation, allowing the coexistence, we find that there is a region in the phase diagram where Kekule order is favored as a result of the competition with the hidden superconducting order, but the latter eventually sets in and coexists with the Kekule state. Fluctuations may play an important role here, and addressing this problem requires the use of sophisticated field-theoretical renormalization group techniques. \cite{RJ,Scherer}
Even though our calculations suggest that if attractive interaction dominates in a self-assembled nanocrystal, it will be of the on-site type, they also indicate that there may be circumstances, as for instance even stronger screening, in which this result could be overturned, so that the nearest-neighbor attraction could take over. This will hopefully motivate further search for materials where this will be the case, and would therefore open up a possibility for the realization of the exotic superconducting states in Dirac materials.

\section{Acknowledgments}

The authors would like to thank L.-K. Lim, I. Herbut, B. Roy and D. Vanmaekelbergh for fruitful discussions. This work is part of the D-ITP consortium, a program of the Netherlands Organisation for Scientific Research (NWO) that is funded by the Dutch Ministry of Education, Culture and Science (OCW). C. M. S. and V. J. acknowledge financial support from the NWO.

\appendix
\begin{widetext}
\section{Effective Hubbard terms} \label{appendixsection integratingoutphonons}

The electron-phonon action for a phonon on sublattice $A$ is given in Eq.~(\ref{eqn electronphononactionforsublatticea}). To integrate out the phonons, the square needs to be completed, which yields
\begin{eqnarray}
&& S_{{\rm el-ph}} \left[\psi^\dagger, \psi; \phi^\dagger, \phi\right]  =  \sum_{{\bf q}, n} \sum_{\sigma, \sigma'} \left(-i\hat{\omega}_{n} + \hbar \omega_{A, E}\right) \left[\phi^{\dagger}_{A, {\bf q}, n} + \frac{1}{\sqrt{\hbar \beta}} \frac{1}{-i\hat{\omega}_{n} + \hbar \omega_{E}} \left(u_0 \, \rho_{A, {\bf q}, \sigma, n} + v({\bf q}) \, \rho_{B, {\bf q}, \sigma, n} \right)  \right] \nonumber \\
&& \times \left[\phi_{A, {\bf q}, n} + \frac{1}{\sqrt{\hbar \beta}} \frac{1}{-i\hat{\omega}_{n} + \hbar \omega_{E}} \left(u_0 \, \rho_{A, -{\bf q}, \sigma', -n} + v(-{\bf q}) \, \rho_{B, -{\bf q}, \sigma', -n}\right) \right] \nonumber \\
&& - \frac{1}{\hbar \beta}  \sum_{{\bf q}, n} \sum_{\sigma, \sigma'} \frac{1}{-i\hat{\omega}_{n} + \hbar \omega_{E}} \left(u_0 \, \rho_{A, {\bf q}, \sigma, n} + v(q) \, \rho_{B, {\bf q}, \sigma, n} \right) \left(u_0 \, \rho_{A, -{\bf q}, \sigma', -n} + v(-{\bf q}) \, \rho_{B, -{\bf q}, \sigma', -n}\right) + A \leftrightarrow B. \nonumber
\end{eqnarray}
 Here, $u_0 \equiv V_0/\sqrt{N}$ and $v({\bf q}) \equiv \tilde{V}_0 \gamma_{{\bf q}} / \sqrt{N}$ with $\gamma_{{\bf k}} \equiv \sum_{\alpha} {\rm e}^{i {\bf k} \cdot \boldsymbol\delta_\alpha}$.
Plugging this expression into the partition function leads to
\begin{eqnarray}\label{eq:effectiveA}
&& \mathcal{Z} = \int \mathcal{D}\left[\psi^{\dagger}, \psi \right] \int  \mathcal{D}\left[\phi^{\dagger}, \phi \right]  {\rm e}^{- \frac{1}{\hbar \beta} S_{{\rm el-ph}} \left[\psi^\dagger, \psi; \phi^\dagger, \phi\right] } = \int \mathcal{D}\left[\psi^{\dagger}, \psi \right] {\rm e}^{- \frac{1}{\hbar \beta} S_{{\rm eff}} \left[\psi^\dagger, \psi \right] }, \nonumber
\end{eqnarray}
where
\begin{eqnarray}
&& S_{\rm eff} \left[\psi^\dagger, \psi\right] = - \frac{1}{\hbar \beta} \sum_{{\bf q}, n} \sum_{\sigma, \sigma'} \frac{\hbar \omega_E}{\hat{\omega}^2_n + \left(\hbar \omega_E\right)^2} \left[ \left(u_0 \rho_{A, {\bf q}, \sigma, n} + v(q) \rho_{B, {\bf q}, \sigma, n} \right) \left(u_0 \rho_{A, -{\bf q}, \sigma', -n} + v(q) \rho_{B, -{\bf q}, \sigma', -n} \right) \right] \nonumber \\
&& - \frac{1}{\hbar \beta} \sum_{{\bf q}, n} \sum_{\sigma, \sigma'} \frac{\hbar \omega_E}{\hat{\omega}^2_n + \left(\hbar \omega_E\right)^2} \left[ \left(u_0 \rho_{B, {\bf q}, \sigma, n} + v(q) \rho_{A, {\bf q}, \sigma, n} \right) \left(u_0 \rho_{B, -{\bf q}, \sigma', -n} + v(q) \rho_{A, -{\bf q}, \sigma', -n} \right) \right].
\end{eqnarray}

 Eqs.\ (\ref{eq:effectiveA}) and (\ref{eqn hubbardhamiltonianinrealspace}) yield effective electron-electron  on-site and on nearest-neighbor interactions in the form
\begin{eqnarray}
&& \tilde{U}({\bf q}) = -  \frac{1}{\hbar \beta} \left\{\frac{U}{N} - 2 \frac{\hbar \omega_{E}}{\hat{\omega}_{n}^{2} + (\hbar \omega_{E})^2} \left[ u_0^2   + v({\bf q}) v(-{\bf q}) \right] \right\}, \nonumber \\
&& \tilde{V}({\bf q}) = - \frac{1}{\hbar \beta} \left\{ \frac{V}{N} \gamma_{\bf q} - 2 \frac{\hbar \omega_{E}}{\hat{\omega}_{n}^{2} + (\hbar \omega_{E})^2} \left[u_0 v(-{\bf q})  + u_0 v({\bf q})\right] \right\}. \nonumber
\end{eqnarray}
Using that at finite temperature the zero Matsubara mode is dominant, and that $|{\bf q}|a \ll 1$ yielding ${\rm exp}\left(i {\bf q} \cdot \boldsymbol\delta_\alpha \right) \approx 1$ such that $\gamma_{{\bf q}} \simeq 3$, we then obtain the results in Eqs.~(\ref{eqn effectivehubbarduinteraction}) and (\ref{eqn effectivehubbardvinteraction}).

\section{Critical Interaction for Hidden Order} \label{appendixsection criticalinteractionhiddenorder}

To obtain the critical interaction for hidden order parameter in Eq.~(\ref{Vc-hidden}), we use
\begin{equation}\label{eq:D1}
\int {\rm d}u \, u^2 {\rm tanh}(u) = \frac{u^3}{3} + u^2 {\rm ln}\left|1 + {\rm e}^{-2u}\right| - u \, {\rm Li}_2 \left(- {\rm e}^{-2u} \right) - \frac{1}{2} {\rm Li}_3 \left(-{\rm e}^{-2u}\right) + C,
\end{equation}
with $C$ as a constant and $Li_n(x)$ is the polylogarithm function of the order $n$.

\section{Solving the Hidden Order Gap Equation} \label{appendixsection solvingthezerotemperaturehiddengapequation}

To solve the hidden order gap equation, we use the following
\begin{eqnarray}
1 &=& \sum_{s = \pm} \int \frac{{\rm d}{\bf q}}{(2 \pi)^2} \frac{|q|^2}{v_F |q| + s \mu} {\rm tanh} \left[\frac{\beta \left(v_F |q| + s \mu\right)}{2} \right] = \frac{1}{v_F^4} \frac{1}{2\pi} \sum_{s = \pm} \int_{s \mu}^{v_F \Lambda} {\rm d}u \frac{\left(u - s \mu \right)^3}{u} {\rm tanh} \left(\frac{\beta_c u }{2}\right) \nonumber \\
&=& \frac{1}{v_F^4} \frac{1}{2\pi} \sum_{s = \pm} \int_{s \mu}^{v_F \Lambda} {\rm d}u \left[ \left(u^2 + 3 \mu^2 \right) - s \mu \frac{\left(3 u^2 + \mu^2 \right)}{u} \right] {\rm tanh} \left(\frac{\beta_c u }{2}\right). \nonumber
\end{eqnarray}
This integral can now be solved using Eq.~(\ref{eq:D1}).

\section{Critical coupling for $s$-Wave Superconductor} \label{appendixsection swavescreviewed}

For the $s$-wave order parameter, we find the following thermodynamical potential\cite{uchoa-2005}
\begin{equation}
\Omega_{\Delta_0} = 4 N \tilde{U} \left|\Delta_0\right|^2 - \frac{1}{\beta} \sum_{{\bf q}; s, s' = \pm} {\rm ln}\left[1 + {\rm exp} \left(- \beta s \sqrt{\left(v_F |q| + s' \mu\right)^2 + \tilde{U}^2 \left|\Delta_0\right|^2} \right) \right]. \nonumber
\end{equation}
Minimizing with respect to the $s$-wave gap $\Delta_0$ leads to the following finite-temperature gap equation
\begin{equation}
1 = \frac{\tilde{U}}{2} \int \frac{{\rm d}{\bf q}}{(2 \pi)^2} \sum_{s = \pm} \frac{1}{\tilde{\omega}_{\Delta_0; s}} {\rm tanh} \left(\frac{\beta \tilde{\omega}_{\Delta_0; s}}{2} \right), \label{SC in CdSe: gap equation for s wave and hidden order with the hidden order turned off for delta not gap for finite temp}
\end{equation}
where
\begin{equation}
\tilde{\omega}_{\Delta_0; s} = \sqrt{\left(v_F |q| + s \mu \right)^2 + \tilde{U}^2 \left|\Delta_0\right|^2}.
\end{equation}
The finite-temperature gap equation corresponds to the one for the Kekule order up to a prefactor, which is due to the gapped nature of both order parameters. Therefore, by minimizing the above thermodynamic potential and setting $T=0$, we obtain the critical interaction for the $s-$wave Dirac superconductor
\begin{equation}
\tilde{U}_c = 2 \pi v_F \left[\Lambda - \frac{2 {\rm ln}(2)}{\beta v_F} \right]^{-1}.
\end{equation}
We see that this critical interaction is smaller than those for the Kekule and hidden order, showing that the $s$-wave superconducting order is preferred. Since both the $s-$wave and Kekule order parameter open up a gap at the Dirac points, the zero-temperature gaps and critical temperature are of the same form, and only differ in the prefactors.

\section{Effective couplings}
\label{appendix_couplings}

The parameters that define the effective interactions in Eqs.~(\ref{eqn effectivehubbarduinteraction}) and (\ref{eqn effectivehubbardvinteraction}) are obtained numerically by calculating the electrostatic interactions between electrons placed on the superlattices. The bare on-site Coulomb interaction of the Hubbard Hamiltonian is given by
\begin{equation}
U \equiv U(\infty) = \int {\rm d}^3 r \, \phi_{A}^{\infty}({\bf r}) \rho_{A}({\bf r}),
\end{equation}
where $\rho_{A}({\bf r})$ is the charge density corresponding to one electron in the $s$ state of a nanocrystal A, and $\phi_{A}^{\infty}({\bf r})$ is the induced potential calculated by solving the Poisson equation using the high-frequency values for the dielectric constants of the inner and outer materials. Similarly, the bare nearest-neighbor Coulomb interaction is given by
\begin{equation}
V \equiv V(\infty) = \int {\rm d}^3 r \, \phi_{A}^{\infty}({\bf r}) \rho_{B}({\bf r}),
\end{equation}
where $\rho_{B}({\bf r})$ is the charge density of an electron placed on a nanocrystal $B$, neighbor of $A$. For reasons that we clarify below, we also calculate $U(0)$ and $V(0)$ using the static dielectric constants instead of the high-frequency ones, thereby including the polarization coming from the LO phonons.

The electron-phonon coupling terms $V_0$ and $\tilde{V}_0$ can be derived by writing the energy of the system in different electrostatic configurations. If we put one electron on a site on sublattice $A$, the classical energy derived from our model Hamiltonian is
\begin{equation}
E(Q) = \frac{\omega_{E}^2 Q^2}{2} + V_0 \sqrt{\frac{2\omega_{E}}{\hbar}}Q,
\end{equation}
where $Q \equiv \sqrt{\hbar/(2\omega_{E})} \left(c^{\dagger}_{A} + c_{A}\right)$ is the operator corresponding to the displacement of the ions in response to the presence of the electron on sublattice $A$. The minimum of $E(Q)$ at $Q_0 = -\sqrt{2}V_{0}/\left(\hbar \omega_{E}^{3/2}\right)$ gives the relaxation energy of the system after injection of the electron in nanocrystal $A$, the so-called Franck-Condon energy $d_{\rm FC} = V_{0}^{2}/(\hbar \omega_{E})$. This energy is also given by the difference $\left[ U(\infty)-U(0) \right]/2$ in the self-energy of the electron on a site on sublattice $A$ in absence and in presence of the ionic response. The factor 1/2 comes from the adiabatic build-up of the charge in nanocrystal $A$. We then find
\begin{equation}
V_0 = \sqrt{\frac{\hbar\omega_{E}}{2} \left[ U(\infty)-U(0) \right]}.
\end{equation}

In order to calculate $\tilde{V}_{0}$, we consider the quantity $U(0)-V(0)$ which can be seen as the energy required to move a test charge (electron) from site $B$ to $A$, when there is already an electron in A which induces the response of the ions. The analog{\color{red}ue} of this quantity derived from our model Hamiltonian is
\begin{equation}
\left[ U(\infty) +2V_{0}\sqrt{\frac{2\omega_{E}}{\hbar}} Q_{0} \right] - \left[ V(\infty) + (V_{0}+\tilde{V}_{0}) \sqrt{\frac{2\omega_{E}}{\hbar}} Q_{0} \right],
\end{equation}
from which we find after some algebra,
\begin{equation}
\tilde{V}_{0} = V_{0} \frac{V(\infty)-V(0)}{U(\infty)-U(0)}.
\end{equation}

The quantities $U(0)$, $U(\infty)$, $V(0)$, and $V(\infty)$ are calculated numerically using the electron wave-functions directly derived from the atomistic tight-binding of Refs.~[\onlinecite{kalesaki, crispaper, 2Dmat}]. The charges on each atom (Cd, Pb, Se) are approximated by point charges, from which we find the potentials by integrating the Poisson equation.

\end{widetext}

\end{document}